\documentclass{article}
\newcommand{\bfr}{\begin{flushright}}
\newcommand{\efr}{\end{flushright}}
 
\begin{document}
\title{Kac-Moody Symmetry in Hosotani Model
}
\author{ 
Kiyoshi Shiraishi\\
Department of Physics, Tokyo Metropolitan University,\\
Setagaya-ku, Tokyo 158, Japan\\
}
\date{Prog. Theor. Phys. {\bf 80} (1988) pp. 601--606,\\ 
Progress Letters
}
\maketitle
\begin{abstract}
The symmetry of the massive tower of fields in higher-dimensional
Yang-Mills theory compactified on a space-time of the form $M_d \times
S^1$ is clarified. The transformations form a loop algebra, a class of
Kac-Moody algebras. Since the symmetry is spontaneously broken, vector
fields ``eat'' Goldstone bosons and acquire masses. The field of
zero-mass mode can also become massive provided that the field of the
internal component develops a vacuum expectation value. The relation
between the ``restoration'' of the symmetry in massive modes and the
gauge transformation of the zero-mode vacuum field is discussed.
\end{abstract}

\bigskip

Several years ago Hosotani \cite{1} offered a symmetry breaking
mechanism which utilizes the gauge field of an internal component. He
considered $SU(N)$ Yang-Mills gauge field and matter fields on $M_3
\times S^1$, where $M_3$ is a three-dimensional space-time and $S^1$ is a
circle.

Recently, Svetvo\v{i} and Khariton \cite{2} used the mechanism in $SU(5)$
gauge theory with matter fields on $M_4 \times S^1$. They attempted to
preserve a large mass hierarchy at gauge symmetry breaking by Hosotani
mechanism.%
\footnote{In a different perspective, the vacuum gauge field on a torus
is proposed as an origin of the feeble ``fifth force''.\cite{3}} 
In the same line, the mechanism has been generalized
and applied to multidimensional unified theories. In such models, it is
said that gauge symmetry is broken by Wilson loops on a general
non-simply connected manifold, whereas supersymmetry remains
unbroken, for example.\cite{4}
 
Hosotani model, in which the non-simply connected space is $S^1$, is the
simplest one, so that we can investigate it for a general Yang-Mills
group \cite{5} and study quantum and/or thermal effects on the model.
\cite{1,2,6}

On the other hand, Dolan and Duff \cite{7} exposed the infinite
parameter symmetries in Kaluza-Klein theories in five dimensions, i.e.,
Einstein gravity theory on a space-time $M_4 \times S^1$. They explained
the masses for excited modes in terms of spontaneous breakdown of the
symmetries. Naturally, we suspect that we can discuss the symmetry
breaking as well as massive tower of vector fields by analysis of the
Yang-Mills theory in higher dimensions in a similar way.

In this paper, we analyze the symmetry of massive fields in Yang-Mills
theory in a space with $M_d\times S^1$. Throughout this paper, We treat
only the classical theory, although a further analysis will provide a
new prospect even in quantum nature of Hosotani model and general
Wilson-loop mechanism.

Let us examine pure Yang-Mills theory in $d+1$ dimensions. The Lagrangian
is
\begin{equation}
 L={\rm tr }\left(-\frac{1}{4}F_{MN}F^{MN}\right)\, , 
\end{equation}
where the field strength $F_{MN}=\partial_MA_N-\partial_NA_M+i[A_M,
A_N]$. Denote the coordinates in $M_d\times S^1$ by $z^M=(x^\mu, y)$,
$\mu=0, 1, 2, \cdots, d-1$.

The Lagrangian is invariant under the infinitesimal gauge
transformation:
\begin{equation}
A_M\rightarrow A'_M=A_M+\Delta A_M=A_M+\nabla_M \Lambda\,,
\label{2}
\end{equation}
where $\nabla_M\Lambda=\partial_M\Lambda+i[A_M, \Lambda]$.
  
The equation of motion is
\begin{equation}
\nabla_MF^{MN}=0\,,
\label{3}
\end{equation}
at classical (tree) level, then vanishing field strength
\begin{equation}
\langle 0|F_{MN}|0\rangle=\langle F_{MN} \rangle=0 \,,
\end{equation}
is a solution to Eq.(\ref{3}).
  
The gauge field can be expressed as a form of Fourier series:
\begin{equation}
A_M(x,y)=\sum_{n=-\infty}^\infty A_{Mn}(x)\,e^{in\theta}\,,
\end{equation}
where $\theta=y/r$, $r$ is the radius of the circle $S^1$. Thus, there
appear infinitely many four-dimensional fields.
  
We also Fourier expand the infinitesimal gauge parameter $\Lambda(z)$ in
the form
\begin{equation}
\Lambda(x,y)=\sum_{k=-\infty}^\infty \Lambda_{k}(x)\,e^{ik\theta}\,.
\end{equation}
Then the infinite number of transformations are parametrized by
$\Lambda_k(x)$. At the same time, the $k$-th transformation on the $n$-th
field is defined by \cite{7}
\begin{equation}
\Delta A_M(x,y)\equiv \sum_{k=-\infty}^\infty\sum_{n=-\infty}^\infty
\Delta_kA_{Mn}(x)\, e^{in\theta}\,,
\end{equation}
and then
\begin{eqnarray}
\Delta_kA_{\mu n}(x)&=&\delta_{nk}\partial_\mu\Lambda_k+
i[A_{\mu (n-k)},\Lambda_k] \,,\\
\Delta_kA_{yn}(x)&=&i\delta_{nk}r^{-1}n\Lambda_k+
i[A_{y (n-k)},\Lambda_k]\,.
\label{8b}
\end{eqnarray}
 
One can soon verify the following commutation relations among these
``gauge'' transformations $\Delta_k(\Lambda_k)$:
\begin{equation}
[\Delta_k(\Lambda_k), \Delta_\ell(\Lambda_\ell)]=
i\Delta_{k+\ell}(\Lambda_{k+\ell})\,,
\end{equation}
where $\Lambda_{k+\ell}=[\Lambda_k,\Lambda_\ell]$, i.e., the algebra of
gauge transformations closes and is just a (Kac-Moody-like) loop
algebra. However, as in the case of the Kaluza-Klein gravity in
Ref.\cite{7}, most of the gauge invariance are spontaneously broken.
  
Suppose that vacuum expectation values of fields are identically zero
except for $\langle 0|A_{y0}|0\rangle$.%
\footnote{The vacuum expectation value of $A_\mu$ breaks the symmetry of
(Lorentz) rotation and may be reduced to zero by continuous gauge
transformations in an ordinary sense.}
That is,
\begin{equation}
\langle 0|A_{y0}|0\rangle=\langle A_{y0}\rangle=\mbox{const}
\end{equation}
is a solution to the equation of motion.
 
From Eq.(\ref{8b}), the expectation value of the infinitesimal variation
is
\begin{equation}
\langle
0|\Delta_kA_{yn}(x)|0\rangle=i\delta_{nk}\{r^{-1}n\Lambda_k+[\langle
A_{y0}\rangle, \Lambda_k]\}\,. 
\label{11}
\end{equation}

If $\langle A_{y0}\rangle=0$, the same explanation as in Ref.\cite{7}
holds. That is to say, for $n\ne 0$, $A_{yn}$ is the Goldstone bosons
associated with the spontaneously broken symmetry, and is absorbed by
vector fields $A_{\mu n}$ with $n\ne 0$; the longitudinal degee of
freedom of massive vector fields comes from these Goldstone bosons. The
field with $n=0$ is still massless provided that
$\langle A_{y0}\rangle=0$.
 
Now, we consider $\langle A_{y0}\rangle\ne 0$ case. This provides a
possible mechanism of mass generation to the zero-mode sector.
 
 To make the analysis simple, we make full use of the residual gauge
freedom in $\langle A_{y0}\rangle$. First, gauge transformation (rigid
``rotation'' in the group space) reduces a general $\langle
A_{y0}\rangle$ to the form,\cite{5}
\begin{equation}
\langle A_{y0}\rangle=\sum_{a'=1}^I  A_y{}^{a'}_0H^{a'}\,,
\label{12}
\end{equation}
where $H^{a'}$ are mutuauy commuting generators of the gauge group $G$,
i.e., they coustitute the Cartan subalgebra of $G$. $I$ is the rank of
$G$. Furthermore, we express the matrix of eigenvalues of $\langle
A_{y0}\rangle$ by use of the weights (for the adjoint representation, it
is equivalent to the roots). Because we treat only gauge field in this
paper, we consider the adjoint representation for a while. The weight
vectors are
\begin{equation}
\lambda_a=(\lambda^1_a,\cdots,\lambda^I_a)\, ,                        
\label{13}
\end{equation}
where $a=1,\cdots, R$. $R$ is the dimension of the adjoint
representation, i.e., dimension of $G$. We use the suffices $a, b,
c,\cdots$  for the adjoint representation, as usual. Then we can define
the following $R \times R$ matrix for later convenience:
\begin{equation}
(\langle A_{y0}\rangle)_{ab}=\sum_{a'}\langle
A_y{}^{a'}_0\rangle\lambda_a^{a'}\delta_{ab} \,.\quad  (\mbox{no sum on
}a)
\label{14}
\end{equation}

The relation to the previous notation is, $A_M=\sum A_M^aT^a$, where
$T^a$ are the generators (which satisfy $[T^a, T^b]= if^{abc}T^c$, where
$f^{abc}$ is the structure constant of $G$). Using (\ref{14}), we can
rewrite (\ref{11}) as
\begin{equation}
\langle
0|\Delta_kA_{y}{}^a_{n}(x)|0\rangle=i\delta_{nk}\{r^{-1}n\Lambda_k^a+
(\langle
A_{y0}\rangle)^{ab} \Lambda_k^b\}\,. 
\label{15}
\end{equation}
where $a=1,\cdots,R$, and we rewrite $\Lambda$ with the use of adjoint
indices.
 
As a concrete example, we examine the group $G=SU(2)$. ($I=1$, and $R=3$
for the adjoint representation.) Then, without loss of generality, we
choose simply, the $3 \times 3$ matrix as follows:
\begin{equation}
(\langle
A_{y0}\rangle)^{ab}\propto \mbox{diag} (1, - 1, 0)\, .                   
\label{16}
\end{equation}

Interestingly enough, two terms on the right-hand side of Eq.(\ref{15})
can be cancelled with each other for $a=1$ and $2$, for some values of
$n=k$.
For instance, if we set
\begin{equation}
(\langle
A_{y0}\rangle)^{ab}=-r^{-1} \mbox{diag} (1, - 1, 0)\,,
\label{17}
\end{equation}
then
\begin{equation}
\langle 0|\Delta_1A_{y}{}^1_{1}(x)|0\rangle=\langle
0|\Delta_{-1}A_{y}{}^2_{-1}(x)|0\rangle=0 \,,
\label{18a}
\end{equation}
but it becomes
\begin{equation}
\langle
0|\Delta_{0}A_{y}{}^1_{0}(x)|0\rangle\ne 0 \quad\mbox{and}\quad \langle
0|\Delta_{0}A_{y}{}^2_{0}(x)|0\rangle\ne 0\,.    
\label{18b}
\end{equation}
As a consequence, the fields $A_\mu{}^1_1$ and $A_\mu{}^2_{-1}$ become
massless, while
$A_\mu{}^1_0$ and $A_\mu{}^2_{0}$ become massive.
  
Here, what is the difference between $\langle
A_{y0}\rangle=0$ and $\langle
A_{y0}\rangle\ne 0$ case? The
same pattern of massive tower is recognized by standard analysis of
mass spectra in both cases \`ala Kaluza-Klein. Indeed, it is already
known that both the cases are mutually related by global gauge
transformations which depend on $\theta$.\cite{1} Let us see how these
transformations are related to the symmetry in the infinite number of
the fields. We restrict ourselves to the case of
$SU(2)$.

We consider a gauge function in the adjoint representation of $SU(2)$ and
show simply what the transformation we consider is. For such a function
the following finite gauge transformation is allowed:
\begin{equation}
\Lambda^a\rightarrow\Lambda'^{a}=\left(\exp i\theta\left(
\begin{array}{lll}
m & & \\
 & -m & \\
 & & 0
\end{array}
\right)\right)^{ab}\Lambda^b \,,
\label{19}
\end{equation}
where $m$ is an integer; thus the transformation is single-valued with
respect to $\theta$. For general gauge groups, the transformation is
\begin{equation}
\Lambda^a\rightarrow\Lambda'^{a}=U^{ab}\Lambda^b
=(\exp -i\theta\sum m^{a'}
\lambda^{a'}_c\delta_{cd})^{ab}\Lambda^b 
\label{20}
\end{equation}
with a constraint
\begin{equation}
U(\theta=2\pi)=1\,.                  
\label{21}
\end{equation}
Therefore $m^{a'}$ are integers when the weight (root) system is properly
normalized.
 
The constraint (\ref{21}) is required for a proper interpretation of the
dimensional reduction, or in other words, for a non-singular gauge
transformation in a usual sense.

Returning to $SU(2)$, according to the transformation matrix (\ref{19}),
the {\em variation} of the gauge field (according to (\ref{2})) is
transformed as follows:
\begin{eqnarray}
\Delta A_\mu^a(z)&\rightarrow&\Delta(\Lambda')A_\mu^a(z)\,,\\
\Delta
A_y^a(z)&\rightarrow&\Delta(\Lambda',\langle
A'_{y0}\rangle)A_y^a(z)\,,
\end{eqnarray}
where $\Delta(\Lambda')$ means the variation which is defined by
(\ref{2}), but with the infinitesimal gauge parameter $\Lambda$ replaced
by
$\Lambda'$, which is obtained by the transformation (\ref{19}). At the
same time, the vacuum gauge field is transformed into the following form:
\begin{equation}
(\langle
A'_{y0}\rangle)^{ab}\rightarrow (\langle
A'_{y0}\rangle)^{ab}- r^{-1}\mbox{diag} (m, -m, 0)\,.
\label{24}
\end{equation}

Then, by a finite rotation of (\ref{19}), the variation of each field is
transformed as follows:%
\footnote{It may be supposed that the shift of the vacuum gauge field
(\ref{24}) compensates the contribution from the transformation of
$\Lambda$.}
\begin{eqnarray}
\Delta_k A_\mu{}_n^1(x)&\rightarrow&\Delta_k
A'_\mu{}_n^1(x)=\Delta_{k-m} A_\mu{}_{n-m}^1(x)\,,\\
\Delta_k A_\mu{}_n^2(x)&\rightarrow&\Delta_k
A'_\mu{}_n^2(x)=\Delta_{k+m} A_\mu{}_{n+m}^2(x)\,,\\
\Delta_k A_\mu{}_n^3(x)&\rightarrow&\Delta_k
A'_\mu{}_n^3(x)=\Delta_{k} A_\mu{}_{n}^3(x)\,,\\
\Delta_k A_y{}_n^1(x)&\rightarrow&\Delta_k
A'_y{}_n^1(x)=\Delta_{k-m} A_y{}_{n-m}^1(x)\nonumber \\
&&=i\delta_{nk}\{r^{-1}(n-m)\Lambda_{k-m}^1+(\langle
A_{y0}\rangle)^{1b}\Lambda_{k-m}^b\}\,,\\
\Delta_k A_y{}_n^2(x)&\rightarrow&\Delta_k
A'_y{}_n^2(x)=\Delta_{k+m} A_y{}_{n+m}^2(x)\nonumber \\
&&=i\delta_{nk}\{r^{-1}(n+m)\Lambda_{k+m}^2+(\langle
A_{y0}\rangle)^{2b}\Lambda_{k+m}^b\}\,,\\
\Delta_k A_y{}_n^3(x)&\rightarrow&\Delta_k
A'_y{}_n^3(x)=\Delta_{k} A_y{}_{n}^3(x)\nonumber \\
&&=i\delta_{nk}\{r^{-1}n\Lambda_{k}^3+(\langle
A_{y0}\rangle)^{3b}\Lambda_{k}^b\}\,.
\end{eqnarray}

Now, we see that the transformation of the type of (\ref{19}) can be
regarded as the change of the ``label'' of the fields. Therefore, the
case we considered through Eqs.(\ref{17}), (\ref{18a}) and (\ref{18b}) is
supplied by assuming the fInite transfomation (\ref{19}) with $m=1$ on
the case with trivial vacuum
$\langle A_y\rangle=0$; two cases are gauge equivalent. At the same time,
we can read the periodicity with respect to the value of the vacuum
field. This fact tells the existence of the periodicity in the (one-loop)
effective potential for Hosotani model. The analysis as shown here will
help understanding of the model with various gauge groups and even in the
higher-dimensional tori.
  
We must care about the inclusion of the matter field. For example, in
the previous case of $SU(2)$, the doublet of $SU(2)$ is transfomled by
(\ref{19}) as follows:
\begin{equation}
\lambda\rightarrow\exp\left(
\begin{array}{cc}
i(m/2)\theta & 0\\
0 &  -i(m/2)\theta
\end{array}
\right)\lambda\,.
\label{27}
\end{equation}
The resulting field is periodic or antiperiodic on $\theta$ with
periodicity
$2\pi$. In general $SU(N)$ gauge theory, $Z_N$ symmetry which is
contained in pure gauge theory is broken by the introduction of the
matter fields.
Therefore the transformation causes $N$ fields with different boundary
conditions from a matter field which belongs to the fundamental
representation. The breakdown of this $Z_N$ symmetry is the origin of
the mass generation of the matter field in Hosotani model.\cite{1,6,8}
Note the vacuum gauge fields of general value generate the same effect
as singular gauge transformations on the gauge fields and/or matter
fields.
The ``twisted'' (i.e., antiperiodic in the extra coordinate) field can
also be expanded as ``twisted'' Fourier series. The examination of the
gauge theory with general twisted fields will be done elsewhere.
 
To summarize: We showed the infinitesimal gauge transformation on the
gauge 
fields in $M_d \times S^1$ can be considered the infinite dimensional
symmetry among the tower of the field of $M_d$. The finite transformation
which depends on the extra coordinate is also regarded as the exchange
of the ``suffices'' of the infinitely many fields.
 
Not only the inclusion of matter fields but the introduction of the
ghost fields \cite{9} are important subjects for investigation of the
quantum and thermal property of Hosotani model and the unification
through higher dimensions. We hope to report the study on these issues
in the near future.

\bigskip

The author would like to thank S. Saito for reading this manuscript.
He also thanks Iwanami F\=ujukai for financial support.


\end{document}